\documentclass[12pt]{iopart}

\usepackage{iopams}  

\usepackage{bm}
\usepackage{graphicx}

\begin{document}

\title[Spinning motion of a deformable self-propelled particle]{Spinning motion of a deformable self-propelled particle in two dimensions}

\author{Mitsusuke Tarama$^{1}$ and Takao Ohta$^{2}$}

\address{Department of Physics, Kyoto University, Kyoto, 606-8502, Japan}
\ead{$^1$tarama@ton.scphys.kyoto-u.ac.jp\\$~~~~~~~^2$takao@scphys.kyoto-u.ac.jp}

\begin{abstract}
We investigate the dynamics of a single deformable self-propelled particle which undergoes a spinning motion in a two-dimensional space. 
Equations of motion are derived from the symmetry argument for three kinds of variables. One is a vector which represents the velocity of the centre of mass. The second is a  traceless symmetric tensor  representing deformation. The third is an antisymmetric tensor for spinning degree of freedom. By numerical simulations, we have obtained variety of dynamical states due to interplay between the  spinning motion and  the deformation.
The bifurcations of these dynamical states  are analyzed by the simplified equations of motion.
\end{abstract}

\maketitle

\section{Introduction}\label{sec:introduction}

Dynamics of self-propelled particles have attracted much attention recently from the view point of nonlinear science and non-equilibrium statistical physics.  In non-biological systems, experiments have been conducted, for example,  for oily droplets \cite{Nagai, Toyota, Ban, Thutupalli, Kitahata} and Janus particles  \cite{Ebbens, Sano} which undergo 
chemical reactions with the molecules in the surrounding media.  Theoretical studies have also been carried out  by computer simulations both for the motion of an individual particle and for the dynamics of interacting particles \cite{Yeomans, Kapral1, Kapral2}. Most of the studies assume that the particles are rigid without shape deformation. However, it is pointed out that, if a particle is soft, its shape is deformed when the migration velocity is increased as has been observed experimentally \cite{Nagai} and analyzed theoretically \cite{Krischer, Baer}. 

In biological systems such as living cells and micro-organisms,  shape deformation plays a central role. In fact, migration of biological objects is induced by shape deformation 
and there are experimental  investigations of the correlation of shape change and cell migration \cite{Keren2, Bosgraaf, Li, Maeda2008,Keren3}. Theoretical studies for cell migration have been developed   \cite{Wada, Ishikawa09, Listeria2,Sasai}. Computer simulations of self-propulsion driven by shape changes have also been available based on artificial deformable systems \cite{Kruse, Yeomans2}. 

Recently we have introduced a model system for deformable self-propelled particles and have studied motion of individual particles and collective dynamics of interacting particles  \cite{OhtaOhkuma, Hiraiwa1,Hiraiwa2,  Ohkuma, Itino}.  Dynamics of a single particle under external forces have also been investigated \cite{Tarama}. This set of model equations has been derived, by an interfacial approach,  from an excitable reaction diffusion system both in two and three dimensions  \cite{OOS,SHO}.

In this paper, we extend our model system to take into account a spinning motion.  There are many experiments of spinning self-propulsion in both  non-biological and biological microscopic systems. 
(i) One is an oily droplet to which
a small piece of  solid soap is attached \cite{Takabatake}. This causes a spinning motion as well as the ordinary translational motion of a droplet.  
(ii) The second example is  
a bacterium {\it Listeria monocytogenes} which causes locomotion  together with spinning (helical) motion by polymerization of actin filaments  \cite{Listeria1,Crenshaw,Listeria3, Chen}. Flagellated bacteria such as {\it Escherichia coli} also exhibits 
a spinning motion by rotating helical filaments \cite{DiLuzio, Goldstein}. In these biological examples, the left-handed and right-handed symmetry is broken because the filament has a specific rotating direction.
(iii) The third is 
an anisotropic doublet composed of paramagnetic colloid particles undergoes a translation motion and rotation near a flat boundary under an oscillatory magnetic field \cite{Sagues}.  The rotating direction follows the rotation of the magnetic field. 
(iv) The fourth is 
a discovery of  spontaneous formation of spiral waves in a Dictyostelium cell, which is coupled with shape deformation, migration and rotation of the cell \cite{Sawai}.

In order to make our model as general as possible, we do not rely on any specific objects but derive the model equations only from the symmetry argument in terms of the antisymmetric tensor variable for spinning motion, the vector variable for the translational motion and the traceless symmetric tensor for elongation of a circular particle. We keep the parity symmetry but allow a spontaneous symmetry braking in our model. 
Numerical simulations have been carried out in two dimensions to obtain variety of dynamical states caused by the interplay between the deformation 
and the  spinning motion.
We have also analyzed theoretically the bifurcations of these dynamical states. The case (ii) above is not considered in the present paper since the angular vector of filament rotation is parallel to the migration velocity and hence the dynamics is essentially three-dimensional.

The organization of this paper is as follows; 
in the next section (Sec.~\ref{sec:model}), we introduce the model equations. 
The results of numerical simulations are described in Sec.~\ref{sec:numerical results}. 
The variety of the dynamical states are obtained as shown  in the dynamical phase diagram.
In Sec.~\ref{sec:analysis of bifurcations}, the theoretical analysis is carried out to derive the bifurcations of these dynamical states. 
Discussion for the results obtained is given in Sec.~\ref{sec:discussion}.

\section{Time-evolution equations}\label{sec:model}
 
We consider a self-propelled particle which changes its shape depending on the migration velocity. The degree of freedom of spinning motion around the centre of mass is also introduced. The basic dynamical variables are the velocity of the centre of mass $v_{i}$, the second-rank traceless symmetric tensor $S_{ij}$ for deformation
and the antisymmetric tensor $\Omega_{ij}$ for spinning motion. By considering possible couplings among these variables and retaining some relevant nonlinear terms, the set of time-evolution equations is given by
\begin{eqnarray}
\frac{d v_{i}}{dt} 
 &=& \gamma v_{i} -v_kv_k v_{i} -a_1 S_{ij}v_{j} -a_2 \Omega_{ij} v_{j} ,
 \label{eq:1.1}\\
\frac{d S_{ij}}{dt}
 &=& -\kappa S_{ij} +b_1 ( v_{i} v_{j} -\frac{v_kv_k}{d} \delta_{ij}) \nonumber\\
 &&+ b_2 ( S_{ik}\Omega_{kj} -\Omega_{ik}S_{kj}) + b_3 \Omega_{ik} S_{kl} \Omega_{lj} ,
 \label{eq:1.2}\\
\frac{d \Omega_{ij}}{d t} 
 &=& -\frac{\partial G}{\partial \Omega_{ij}} + c_1 ( S_{ik}\Omega_{kj} +\Omega_{ik}S_{kj}) + 4c_2 S_{ik} \Omega_{kl} S_{lj} ,
 \label{eq:1.3}
\end{eqnarray}
where 
\begin{equation}
G \equiv \frac{\zeta}{2} \tr \Omega^2 +\frac{1}{4} \tr \Omega^4  ,
 \label{eq:1.4}
\end{equation}
and $a_1$, $a_2$, $b_1$, $b_2$, $b_3$, $c_1$ and $c_2$ are the coupling constants and $d$ is the dimensionality of space. Hereafter we will put $d=2$. The repeated indices imply summation. The coefficient $\gamma$ takes either negative or positive values. 
Throughout the present paper, the parameters  $\kappa$ and $\zeta$ are assumed to be positive. The second rank traceless symmetric tensor $S_{ij}$, which represents an elliptical deformation of a circular particle, is defined by 
\begin{equation}
S_{ij} = 
\frac{s}{2}
\left[
\begin{array}{cc}
\cos 2\theta & \sin 2\theta \\
\sin 2\theta & -\cos 2\theta
\end{array}
\right]  ,
 \label{eq:1.6}
\end{equation}
where $s>0$ is the magnitude of deformation and the angle $\theta$ represents the direction of elongation. 
The antisymmetric traceless tensor is defined as 
\begin{equation}
\Omega_{ij} 
 = 
 \left[
 \begin{array}{cc}
 0 & \omega \\
 -\omega & 0
 \end{array}
 \right]
 \label{eq:1.5}
\end{equation}
with $\omega$ a dynamical variable.

By putting $(v_1, v_2)=v(\cos\phi, \sin\phi)$, Eqs.~(\ref{eq:1.1}) - (\ref{eq:1.3}) are written as 
\begin{eqnarray}
\frac{d v}{d t} 
 = \gamma v -v^3 -\frac{a_1}{2} s v \cos 2\psi  ,
 \label{eq:1.7}\\
\frac{d \phi}{d t} 
 = -\frac{a_1}{2} s \sin 2\psi +a_2 \omega  ,
 \label{eq:1.8}\\
\frac{d s}{d t} 
 = -\kappa s +b_1 v^2 \cos2\psi +b_3 s \omega^2  ,
 \label{eq:1.9}\\
\frac{d \theta}{d t} 
 = -\frac{b_1}{2 s} v^2 \sin 2\psi +b_2 \omega  ,
 \label{eq:1.10}\\
\frac{d \omega}{d t} 
 = \left( \zeta -c_2 s^2 \right) \omega -\omega^3   ,
 \label{eq:1.11}
\end{eqnarray}
where $\psi=\theta-\phi$.
Note that the second term on the right hand side of Eq.~(\ref{eq:1.3}) vanishes in two dimensions. 
Because of the isotropy of space, only the relative angle  $\psi$ enters into the set of equations as an independent variable.
In fact, we have from Eqs.~(\ref{eq:1.8}) and (\ref{eq:1.10}) 
\begin{eqnarray}
\frac{d \psi}{d t} 
 = -(\frac{b_1}{2 s} v^2 -\frac{a_1}{2} s) \sin 2\psi +(b_2-a_2) \omega  .
 \label{eq:eqpsi}
\end{eqnarray}

If the anti-symmetric tensor $\Omega$ is not considered, the set of equations (\ref{eq:1.1}) and  (\ref{eq:1.2}) has been studied both in two and three dimensions \cite{OhtaOhkuma, Hiraiwa1, Hiraiwa2}. Furthermore, those equations have been derived from a reaction-diffusion system \cite{OOS,SHO}. See also Ref. \cite{Tarama}  where dynamics of a self-propelled soft particle under external fields have been investigated. When the spinning degree of freedom is absent, i.e.,  $a_2=b_2=b_3=0$, Eqs.~(\ref{eq:1.7}) - (\ref{eq:1.10}) exhibit a bifurcation at $\gamma=\gamma_c$ where
\begin{equation}
\gamma_c = \frac{\kappa}{2} \frac{1+B}{B}
 \label{eq:gamma_c}
\end{equation}
with
\begin{equation}
B = \frac{a_1 b_1}{2 \kappa}   .
 \label{eq:B}
\end{equation}
When $\gamma < \gamma_c$, a straight motion of a particle is stable. However, this motion becomes unstable for $\gamma>\gamma_c$ and a circular motion (orbital revolution) appears. 
It should be noted that the coefficient in front of $\sin 2\psi$ in Eq. (\ref{eq:eqpsi}) vanishes at $\gamma=\gamma_c$.
Another important parameter is the coefficient $b_1$ in Eq.  (\ref{eq:1.2}). When $b_1$ is positive, a particle tends to elongate along the propagating direction whereas when it is negative, it deforms perpendicularly to the translational velocity \cite{OhtaOhkuma}. Actually we have found that $b_1<0$ for the excitable reaction-diffusion system \cite{OOS,SHO}. Throughout this paper, we will put $b_1<0$ and $a_1<0$ so that the constant $B$ defined by Eq. (\ref{eq:B}) is positive provided that the relaxation rate of deformation $\kappa$ is positive.

Before closing this section, we summarize the meaning of other terms in the time-evolution equations (\ref{eq:1.1}), (\ref{eq:1.2}) and (\ref{eq:1.3}). It is evident from Eq.  (\ref{eq:1.8}) that the term with the coefficient $a_2$ tends to curve the trajectory of migration. We choose $a_2>0$ so that a particle spinning to the counter-clockwise direction (viewed from the top) rotates to the left (viewed from behind).  
This is similar to the Magnus force \cite{Bachelor} which is proportional to the density of the surrounding fluid and the cross section of a rigid cylinder and the volume of a rigid sphere.
The $b_2$-term in Eq.  (\ref{eq:1.2}) is the same as the convective term for the orientational tensor in liquid crystals under rotational flow \cite{Gennes} so that we put $b_2=1$. In other words, the $b_2$-term is not dissipative whereas all other terms on the right hand side of Eq. (\ref{eq:1.2}) are dissipative. The last term in Eq.  (\ref{eq:1.2}) with $b_3>0$ represents that spinning motion  enhances deformation whereas the $c_2$-term in Eq.   (\ref{eq:1.3}) with $c_2>0$ has an effect such that an elongated particle prevents from spinning as can be seen from Eq.  (\ref{eq:1.11}). The $c_1$-term drops out in two dimensions. The coefficient $\zeta$ in Eqs. (\ref{eq:1.4}) is chosen to be positive. This means that the particle has an internal mechanism to keep spinning motion. Furthermore, the potential $G$ takes a form that both clockwise and counter-clockwise motions are equally possible. It is noted that the coefficients in the $v^3$ term in Eqs.  (\ref{eq:1.1}) and the $\tr \Omega^4 $ term in (\ref{eq:1.4}) can be eliminated, without loss of generality,  by absorbing those in the definitions of $v_i$ and $\omega$. Quite recently, we have derived Eq. (\ref{eq:1.1}) with $a_1=a_2=0$ in fluids taking into account the Marangoni effect and the hydrodynamic interaction \cite{Yabunaka}. However, such a study for shape deformation and spinning motion has not been available at present. 

\begin{figure}
  \begin{center}
  \includegraphics[width=9cm]{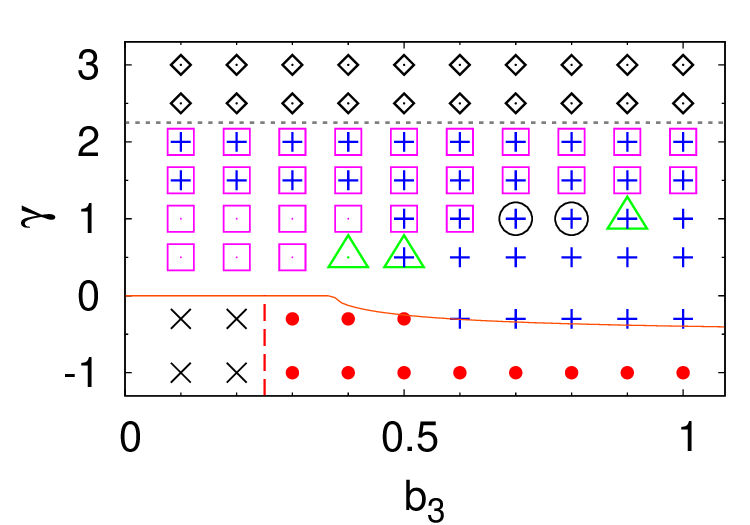} 
      \caption{(color online) 
      Dynamical phase diagram obtained by solving Eqs.~(\ref{eq:1.7}) - (\ref{eq:1.11}) numerically. 
      The  meaning of the symbols is given in the text.
          The bifurcation boundary from the motionless state to the spinning state without migration is indicated by the broken line which is obtained from Eq.~(\ref{eq:b3*}). 
      The thin solid line obtained from Eq.~(\ref{eq:3.53}) represents the boundary between the $v=0$ and the $v\ne0$ region.
      The dotted line given by Eq.~(\ref{eq:3.37}) is the threshold at which  the state $\omega=0$ becomes unstable. 
      The parameters are $\kappa=0.50$, $a_1=-1$, $a_2=0.75$, $b_1=-0.5$, $b_2=1$, $\zeta=2$, and $c_2=2$. 
The bifurcation threshold $\gamma_c$ defined by  (\ref{eq:gamma_c}) is given by $\gamma_c=0.75$.
      } \label{fig:diagram}
  \end{center}
\end{figure}

\section{Numerical Results}\label{sec:numerical results}

In the preceding section, we have fixed the sign of the coefficients in the time-evolution equations  (\ref{eq:1.1}), (\ref{eq:1.2}) and (\ref{eq:1.3}) with  (\ref{eq:1.4}) by considering the mechanisms of each term. However, it is difficult to determine the magnitude of those coefficients within the present phenomenological approach. We are concerned with the effects of spinning on the self-propelled motion. Therefore, we choose the coefficient $\gamma$ in Eq.  (\ref{eq:1.1}) as an important basic parameter since three different kinds of solutions, motionless state, straight motion and circular motion, are realized by changing $\gamma$ in the absence of spinning. The other aspect to be considered is the coupling between spinning and deformation. In Eq.  (\ref{eq:1.2}) for the deformation tensor, the coefficient $b_2$ has been uniquely put to be $b_2=1$. Therefore the remaining coefficient $b_3$ of the nonlinear coupling between $S$ and $\Omega$ is chosen as another basic parameter. Other remaining coefficients are set to be of the order of unity as $\kappa=0.5$, $a_1=-1$, $a_2=0.75$, $b_1=-0.5$, $b_2=1$, $\zeta=2$, and $c_2=2$. 

Figure~\ref{fig:diagram} show the dynamical phase diagram obtained from Eqs.~(\ref{eq:1.7}) - (\ref{eq:1.11}).
In numerical simulations, we have employed the fourth Runge-Kutta method with the time increment $\delta t = 10^{-4}$. 
 The different symbols correspond to different dynamical states as explained in detail below.

\begin{figure}
  \begin{center}
 \includegraphics[width=10cm]{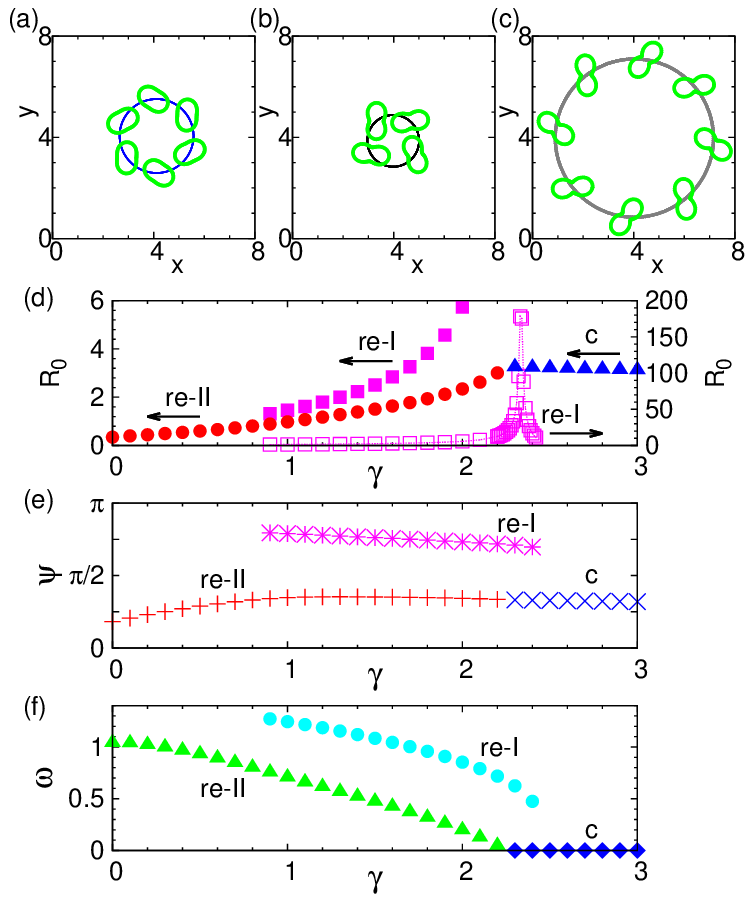} 
      \caption{(color online) 
      Rotation and orbital revolution of a particle in the counter-clockwise direction. Trajectory in real space is displayed for (a)  revolution I motion for $\gamma=1$ and $b_3=0.5$,  (b) revolution II motion for  $\gamma=1$ and $b_3=0.9$,  and (c) circular motion  for $\gamma=3$ and $b_3=0.5$.  The particle size in (a) - (c)  is reduced by the factor of $1/4$ for the sake of clarity. 
   (d) Radius of the circular orbit in real space, (e) the relative angle $\psi$, and (f) the spinning variable
   $\omega$ as a function of  $\gamma$ for  $b_3=0.5$. 
       }
        \label{fig:trace_c}
  \end{center}
\end{figure}

The state indicated by the crosses  in Fig.~\ref{fig:diagram}  is the motionless state. 
In this state,   $v=0$ and $s=0$ and the value of $\omega$ is finite.
That is, the particle with a circular shape is spinning and its centre of mass does not move. 
In the region indicated by the filled circles, the variables $s$ and $\omega$ take finite constant values while the magnitude of the velocity is zero, $v=0$. 
The angle of the elongated direction $\theta$ varies monotonically in time. 
Therefore, the particle is elliptically deformed while the centre of mass does not move and undergoes  clockwise or counter-clockwise rotations depending on the initial condition. 
We call this motion as  a spinning motion throughout this article. 
The white square symbols and the plus symbols mean two different circular motions, which we call the revolution I and  revolution II states, respectively. 
In both of these states, all of the variable $\omega$, $s$, and $v$ take finite constant values and the angles $\phi$ and $\theta$ varies monotonically in time, keeping their difference $\psi=\theta-\phi$ constant in time. 
The trajectories of the revolution I and II motions in real space are displayed in Figs.~\ref{fig:trace_c}(a) and (b), where some snapshots of the particle which undergoes counter-clockwise rotation are shown. 
In both states, clockwise rotation can also appear depending on the initial condition. 
It is noted in Fig.~\ref{fig:diagram} that there is a region where the revolution I and  revolution II states coexist. 
The properties  of the revolution I and II states will be discussed in detail in Sec.~\ref{sec:analysis of bifurcations}.

There is another orbital revolution which appears for large values of $\gamma$ in the region indicated by the diamonds in Fig.~\ref{fig:diagram}.
We call this state as a circular state where  $\omega =0$ while the values of $v$ and $s$ are finite constants and the angle of the propagating direction $\phi$ and the elongation direction $\theta$ varies monotonically with a fixed finite difference $\psi=\theta-\phi$. 
The trajectory and the particle shape of a counter-clockwise circular motion 
 in real space are displayed in Fig.~\ref{fig:trace_c}(c).  
 Since no spinning motion occurs, this is nothing but  the circular motion obtained in Ref.~\cite{OhtaOhkuma}.

The quantitative differences of these three motions are summarized in Figs.~\ref{fig:trace_c}(d), (e) and (f). 
In Fig.~\ref{fig:trace_c}(d), the radius of the circular orbit in real space of the revolution I motion, the revolution II motion, and the circular motion is plotted as a function of  $\gamma$ by the squares, the  circles, and the  triangles respectively.
In drawing a particle, the radius of an undeformed circular particle is set to be $r_0=2$ throughout the present paper.
The radius of the revolution I motion in a larger scale is also plotted by the open squares.  Here, note that the radius of the revolution I motion becomes extremely large around $\gamma=2.33$. 
This will be analyzed theoretically in section \ref{sec:analysis of bifurcations}.
In Figs.~\ref{fig:trace_c}(e) and (f), the values of the variables $\psi$ and $\omega$ are plotted 
for $\omega\ge0$,  respectively. Note that $\psi$ for the revolution I motion (the stars) is larger than $\pi/2$ whereas $\psi$ for the revolution II (the pluses) and circular (the crosses) motions is less than $\pi/2$, which is consistent with the particle configuration in Fig.~\ref{fig:trace_c}(a).
The values of $\psi$ of the revolution II state smoothly connect with those of the circular state at around $\gamma=2.3$. There is no such a connection with that of the revolution I state. 
Similarly, in Fig.~\ref{fig:trace_c}(f),  the values of $\omega$ of the revolution II state (the triangles) continuously connect with those of the circular motion (the diamonds) at around $\gamma=2.3$,  but no such a connection occurs with the revolution I state (the circles).

\begin{figure}
  \begin{center}
  \includegraphics[width=10cm]{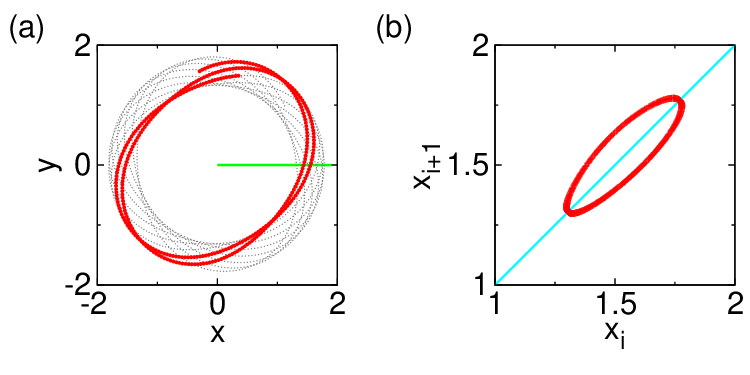} 
      \caption{(color online) 
     (a) Trajectory and (b) return map of the quasi-periodic I motion for $\gamma=1$ and $b_3=0.8$  in the real space.
     The trajectory for a shorter  time interval is indicated by the thick solid line in (a).
           The return map  is obtained at the Poincar\'{e} section indicated  by the horizontal line in (a). 
     } \label{fig:qp1}
  \end{center}
\end{figure}

\begin{figure}[t]
  \begin{center}
  \includegraphics[width=10cm]{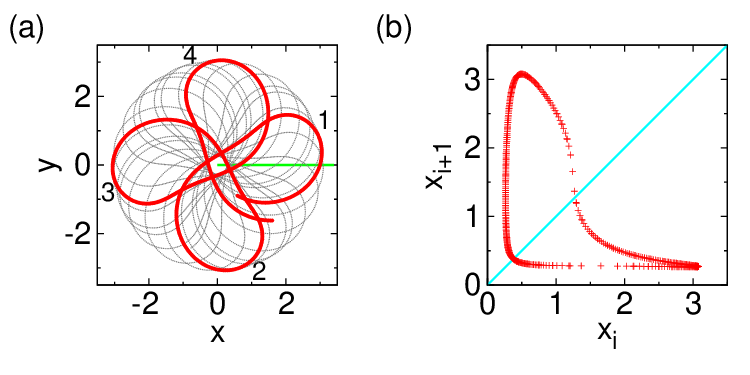} 
      \caption{(color online) 
      (a) Trajectory and (b) return map of the quasi-periodic I motion for $\gamma=1$ and $b_3=0.9$  in the real space.
           The trajectory for a shorter time interval is shown by the thick solid line in (a) together with the chronological order 1, 2, 3,  and 4.
            The return map is obtained at the Poincar\'{e} section indicated by the horizontal line in (a). 
      } \label{fig:qp2}
  \end{center}
\end{figure}

In the region of the  white circles around $\gamma=1$ and $b_3=0.75$  in Fig.~\ref{fig:diagram}, a quasi-periodic I state appears, whose trajectory of a counter-clockwise rotation is displayed in Fig.~\ref{fig:qp1}(a). 
In Fig.~\ref{fig:qp1}(b), we show the return map of this motion, which is obtained at the Poincar\'{e} section indicated by the horizontal line in Fig.~\ref{fig:qp1}(a). 
In this state, all of the variables $v$, $\phi$, $s$, $\theta$, and $\omega$, as well as the difference $\psi=\theta-\phi$ are time-dependent. 
There is another quasi-periodic motion called a quasi-periodic II state in the region of the triangles in Fig.~\ref{fig:diagram}. One example of the trajectory in real space is shown in Fig.~\ref{fig:qp2}(a)  for $\gamma=1$ and $b_3=0.9$,
where a particle rotates in the counter-clockwise direction. Some parts of the trajectory are highlighted by the thick solid line together with the number indicating the chronological order. 
In this state, all of the variables $\omega$ and $v$ and $s$ are non-zero, and both of $\phi$ and $\theta$ as well as  their difference $\psi=\theta-\phi$ are time-dependent. 
In Fig.~\ref{fig:qp2}(b), we show the return map of the quasi-periodic II motion, which is obtained at the Poincar\'{e} section at the horizontal line in Fig.~\ref{fig:qp2}(a). 
Note that both the quasi-periodic  I and II states coexist with the revolution II state (plus symbols) in Fig.~\ref{fig:diagram}.

\begin{figure}[t]
  \begin{center}
   \includegraphics[width=10cm]{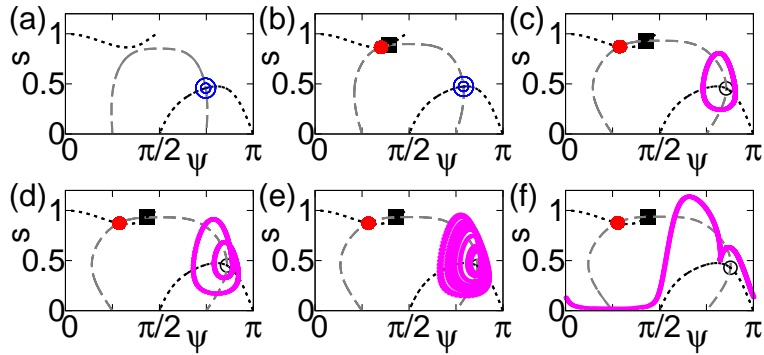} 
      \caption{(color online) 
      Attractors in the $s$-$\psi$ plane for the revolution I and II states, the quasi-periodic I and II states, the period-doubling state, and the chaotic state for positive values of $\omega$ and for $\gamma=1$ and (a) $b_3=0.3$, (b) $b_3=0.5$, (c) $b_3=0.8$, (d) $b_3=0.82$, (e) $b_3=0.844$, and (f) $b_3=0.9$. 
      The broken line is  the nullcline for $ds/dt=0$ 
       and the dotted line is the nullcline  for $d\psi/dt=0$ 
      obtained from the reduced equations (\ref{eq:3.16}) and (\ref{eq:3.18}). The double circle (white circle) is the stable (unstable) fixed point corresponding to the revolution I state.  The black square indicates the saddle point  and the filled circle indicates the stable fixed point corresponding to the revolution II state. 
      The limit cycles in (c) and (f) correspond to the quasi-periodic I and II states respectively. The attractors in (d) and (e) correspond to the period-doubling state and the chaotic state respectively. 
               }
       \label{fig:s_psi}
  \end{center}
\end{figure}

\begin{figure}[t]
  \begin{center}
   \includegraphics[width=10cm]{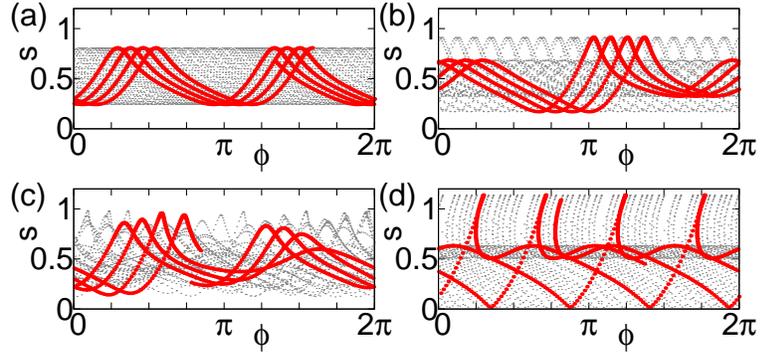} 
       \caption{(color online) 
       Trajectory in the $s$-$\phi$ plane for (a) the quasi-periodic I state, (b) the period-doubling state, (c) the chaotic state, and (d) the quasi-periodic II state.  The thicker (dotted) lines indicate the trajectories for a shorter (longer) time interval.
       The parameters are the same as those in Fig.~\ref{fig:s_psi}(c) - (f) respectively. 
       }
       \label{fig:s_phi}
  \end{center}
\end{figure}

In order to elucidate  further each motion described above, we consider the dynamics on the  $s$-$\psi$ plane. Figure \ref{fig:s_psi} shows  the attractors for $\gamma=1$ and for different values of $b_3$.  In Figs.~\ref{fig:s_psi} (a) for $b_3=0.3$ and (b) for $b_3=0.5$, the stable fixed point indicated by the double circle is the solution of the revolution I state. In Fig.~\ref{fig:s_psi}(b) another stable fixed point of the revolution II state indicated by the black circle appears as a saddle-node bifurcation.
In Fig.~\ref{fig:s_psi}(c) for $b_3=0.8$, the fixed point corresponding to the revolution I state becomes unstable via  a Hopf bifurcation and a limit cycle oscillation appears which represents the quasi-periodic I state. 
Increasing $b_3$, the simple limit cycle of the quasi-periodic I state becomes unstable and a period-doubling occurs as shown in Fig.~\ref{fig:s_psi}(d)  for $b_3=0.82$. By increasing $b_3$ further, a chaotic state appears as shown in Fig.~\ref{fig:s_psi}(e)  for $b_3=0.844$. When $b_3$ is extremely  large, it turns out that $\psi$ varies from 0 to $\pi$  as shown in Fig.~\ref{fig:s_psi}(f)  for $b_3=0.9$, which corresponds to the quasi-periodic II state.
Figure \ref{fig:s_psi} shows the case of positive values of $\omega$. The attractors and the nullclines for negative values of $\omega$ can be obtained by replacing  $\psi$ by $\pi-\psi$.

Now a question arises. Why does  a limit cycle oscillation in the  $s$-$\psi$ plane cause the quasi-periodic I motion in the real space? Similar problems exist for the period-doubling too. 
The origin can be traced back to the isotropy of space. In fact, the dynamics of the system is governed by the set of Eqs.~(\ref{eq:1.7}), (\ref{eq:1.9}), (\ref{eq:1.11}) and (\ref{eq:eqpsi}) whose solutions behave as  shown in Fig.~\ref{fig:s_psi}. However, when we plot the trajectory of the particle in the real space, we have to solve Eq.~(\ref{eq:1.8}) for the angle $\phi$ of the velocity. The magnitude $s$ of deformation is shown in Fig.~\ref{fig:s_phi} as a function of $\phi$ for the parameters corresponding to those of  Fig.~\ref{fig:s_psi}(c) - (f). It is evident from Figs.~\ref{fig:s_psi}(a) and \ref{fig:s_phi}(a) that the trajectory on the $s-\psi-\phi$ space composes a torus and hence a quasi-periodic I motion can appear. Other complex dynamics of 
the period doubling, chaotic motion and quasi-periodic II motion can also be understood from Figs.~\ref{fig:s_phi}(b), (c) and (d), respectively.

\begin{figure}[t]
  \begin{center}
   \includegraphics[width=10cm]{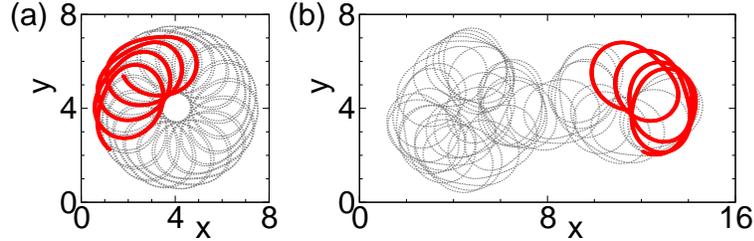} 
      \caption{(color online) 
      Trajectory in the real space of (a) period-doubling state and (b) chaotic state. 
      The trajectory for  a shorter  time interval is shown by the thick solid line.
      The parameters are chosen as $\gamma=1$ and (a) $b_3=0.82$ and (b) $b_3=0.844$. 
      } \label{fig:trace_pd}
  \end{center}
\end{figure}

\begin{figure}[t]
  \begin{center}
    \includegraphics[width=10cm]{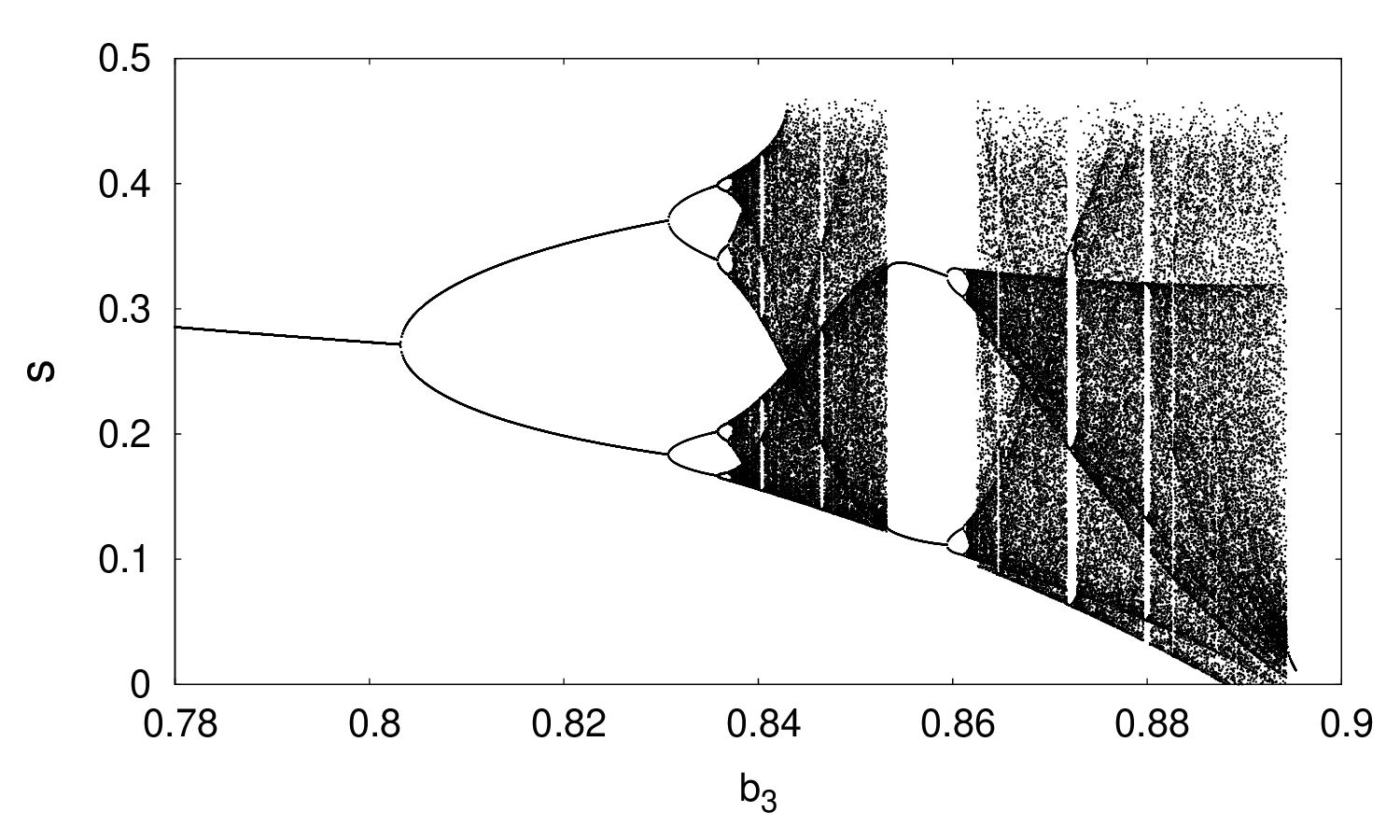} 
      \caption{
      Bifurcation diagram near and in the chaotic regime obtained from Eqs.~(\ref{eq:1.7}) - (\ref{eq:1.11}) with $\gamma=1$ imposing the conditions $\cos2\psi=0.7$, $\sin2\psi<0$, $d\psi/dt>0$, and $\omega>0$. 
      } \label{fig:logistic}
  \end{center}
\end{figure}

As explained above,
the limit cycle corresponding to the quasi-periodic I state becomes unstable by increasing the values of $b_3$ and the period-doubling bifurcation occurs as shown in Fig.~\ref{fig:s_psi}(d). 
The corresponding trajectory in the real space is displayed in Fig.~\ref{fig:trace_pd}(a).
Further increase of $b_3$ leads to a chaotic motion as shown in Fig.~\ref{fig:trace_pd}(b).
This chaotic state eventually disappears for larger values of $b_3$ and, in turn,  the quasi-periodic II state appears.
This series of the dynamical transitions from the quasi-periodic I state  for $b_3=0.8$ to the quasi-periodic II state  for $b_3=0.9$ are not shown  in Fig.~\ref{fig:diagram} due to the size of the grid. However we show,  in Fig.~\ref{fig:logistic},  the bifurcation diagram in the vicinity of the chaotic regime is displayed, which  has been obtained numerically from Eqs.~(\ref{eq:1.7}) - (\ref{eq:1.11}) with $\gamma=1$ under the conditions  $\cos2\psi=0.7$, $\sin2\psi<0$, $d\psi/dt>0$, and $\omega>0$.  This is a typical route to chaos via period-doubling bifurcation well known in a logistic map \cite{logistic}.
Because of the choice $d\psi/dt>0$, the quasi-periodic II state with $\omega>0$, where $\psi$ decreases monotonically, does not appear in Fig.~\ref{fig:logistic}.

\section{Analysis of bifurcations}\label{sec:analysis of bifurcations}

In this section, we analyze the bifurcations found numerically in Section \ref{sec:numerical results}. 
However, direct analysis of set of Eqs.~(\ref{eq:1.7}) - (\ref{eq:eqpsi}) for the four independent variables seems  impossible. Therefore we have to introduce some approximations or simplification. Since our main concern is the coupling between spinning motion and deformation, we retain the deformation degrees of freedom $S$ and eliminate the variables $v$ and $\omega$ by putting $dv/dt=d/\omega/dt=0$ in Eqs.~(\ref{eq:1.7}) and (\ref{eq:1.11}), respectively. The stability and bifurcations of the solutions to the set of Eqs.~(\ref{eq:1.9}) and  (\ref{eq:eqpsi})  are investigated. It is emphasized that this approach is justified when the particle is sufficiently soft and near the stability threshold where a rectilinear motion becomes unstable in the absence of spinning, that is,  for $\kappa \ll 1$ and $\gamma \sim \gamma_c$ where $\gamma_c$ is given by Eq. (\ref{eq:gamma_c}). 

Numerical simulations in the preceding section have been carried out in the condition that all the parameters are of the order of unity. Therefore the comparison with the predictions obtained by the reduced set of Eqs.~(\ref{eq:1.9}) and  (\ref{eq:eqpsi}) is expected to be qualitative. Nevertheless, it will be found that some of the phase boundaries in the dynamical phase diagram in Fig. \ref{fig:diagram} agree semi-quantitatively with the theoretical results. 

By putting $dv/dt=0$ in Eq.~(\ref{eq:1.7}), the velocity $v$ can be expressed as 
\begin{equation}
v = 
\left\{
 \begin{array}{cc}
 0 & ~~\rm{when~} \Gamma \le 0 \\
 \Gamma^{1/2} & ~~\rm{when~} \Gamma > 0
 \end{array}
\right.  ,
 \label{eq:3.1}
\end{equation}
where 
\begin{equation}
\Gamma =  \gamma-\frac{a_1}{2}s \cos2\psi   .
 \label{eq:3.2}
\end{equation}
Equation~(\ref{eq:1.11}) with  $d \omega /dt=0$  gives us three stable solutions:
\begin{equation}
\omega = 
\left\{
 \begin{array}{cc}
 0 & ~~\rm{for~} \zeta-c_2 s^2 \le 0 \\
 \pm \tilde{\omega}(s) & ~~\rm{for~} \zeta-c_2 s^2 >0
 \end{array}
\right.  ,
 \label{eq:3.19}
\end{equation}
where 
\begin{equation}
\tilde{\omega}(s)\equiv \left( \zeta-c_2s^2 \right)^{1/2}   .
\label{eq:omega}
\end{equation}

First, we consider  the motionless state $v=0$.
In this case, the angle of the propagating direction $\phi$ has no meaning. 
Therefore, Eqs.~(\ref{eq:1.9})  and  (\ref{eq:1.10}) become
\begin{eqnarray}
\frac{d s}{d t}
 &=& -\kappa s +b_3 s \omega^2  ,
 \label{eq:3.3}\\
\frac{d \theta}{d t} 
 &=& b_2 \omega  .
 \label{eq:3.4}
\end{eqnarray}
Since $\omega$ is given by Eq.~(\ref{eq:3.19}), Eq.~(\ref{eq:3.3}) is closed for $s$. 
Therefore, it is sufficient to analyze the stability of the fixed point of Eq.~(\ref{eq:3.3}). 
If $\omega=0$, Eq.~(\ref{eq:3.3}) becomes, 
\begin{equation}
\frac{ds}{dt} = -\kappa s   .
 \label{eq:3.42}
\end{equation}
Since $\kappa>0$, we have the solution $s=0$. However, this gives rise to  $\zeta-c_2s^2=\zeta>0$, which
 contradicts the condition $\zeta-c_2s^2<0$ for $\omega=0$ in Eq.~(\ref{eq:3.19}). On the other hand, if $\omega=\pm\tilde{\omega}$, Eq.~(\ref{eq:3.3}) becomes 
\begin{equation}
\frac{ds}{dt} = -b_3 c_2 \left( s^2 -K \right) s   ,
 \label{eq:3.43}
\end{equation}
where we have defined 
\begin{eqnarray}
K = \frac{\zeta-\tilde{\kappa}}{c_2}  ,
 \label{eq:K}\\
\tilde{\kappa} = \frac{\kappa}{b_3}  .
 \label{eq:kappa_tilde}
\end{eqnarray}
We obtain a stable solution $s=0$ for $K<0$,  
which loses its stability at $K=0$ 
and a pair of other stable solutions $s=\pm\sqrt{K}$ 
appears via a pitchfork bifurcation. 
The stability condition $\zeta-c_2s^2>0$ of $\omega=\pm\tilde{\omega}$ in Eq.~(\ref{eq:3.1}) 
 is satisfied for both of the stable solutions. 
The rotating frequency $\omega$ is given from Eq.~(\ref{eq:omega}) by $\pm\sqrt{\zeta}$ for $s=0$  and $\pm\sqrt{\tilde{\kappa}}$ for $s=\pm\sqrt{K}$, respectively.

We identify these stable fixed points with the states with $v=0$ obtained from the original equations (\ref{eq:1.7}) - (\ref{eq:1.11}). 
First, note that, when $s=0$, the circular particle is not deformed  and hence the angle of the longitudinal axis of the deformation $\theta$ loses its meaning. 
Therefore, the solution $(s,\omega)=(0,\pm\sqrt{\zeta})$ corresponds to the motionless state. 
In the same way, the pair of the solutions $(\sqrt{K},\pm\sqrt{\tilde{\kappa}})$ corresponds to the counter-clockwise and clockwise spinning motion. 
The pitchfork bifurcation mentioned above is a bifurcation from the motionless state to the spinning state.  The bifurcation threshold $b_3=b_3^*$ is given by the condition $K=0$ as
\begin{equation}
b_3^* = \frac{\kappa}{\zeta}  .
 \label{eq:b3*}
\end{equation}
The bifurcation threshold (\ref{eq:b3*}) is indicated by the thick broken line in Fig.~\ref{fig:diagram} in a good agreement with the numerical results obtained from Eqs.~(\ref{eq:1.7}) - (\ref{eq:1.11}). 

The condition $\Gamma \le 0$ for $v=0$ in Eq.~(\ref{eq:3.1}) should be satisfied for both the motionless state and the spinning state without migration.
In the motionless state, which is represented by the stable fixed point $(s,\omega)=(0,\pm\sqrt{\zeta})$, the condition becomes $\Gamma=\gamma<0$. 
On the other hand, the derivation of  the condition for the spinning state in terms of the original parameters  is more involved. 
As we have shown above, the variables are given by $(s,\omega)=(\sqrt{K},\pm\sqrt{\tilde{\kappa}})$ in the spinning state.
By substituting  these variables into Eq.~(\ref{eq:3.2}), the condition in Eq.~(\ref{eq:3.1}) is written as 
\begin{equation}
\Gamma = \gamma-\frac{a_1}{2} K^{1/2} \cos 2\psi < 0  .
 \label{eq:3.50}
\end{equation}
Although the variable $\psi$ loses its meaning when $v=0$, we may define it around the bifurcation from $v=0$ to $v \ne 0$. In fact, from Eq.~(\ref{eq:eqpsi}) with $v=0$, we obtain
\begin{equation}
\frac{d \psi}{dt} = \frac{a_1}{2} K^{1/2} \sin 2\psi +\chi_{\omega}\left( b_2 -a_2 \right) \tilde{\kappa}^{1/2}  ,
 \label{eq:3.51}
\end{equation}
where $\chi_{\omega}$ is the sign of $\omega$. 
The steady solution is given by
\begin{equation}
\sin 2\psi = -\frac{2 \chi_{\omega} (b_2-a_2)}{a_1} \left( \frac{\tilde{\kappa}}{K} \right)^{1/2}   .
 \label{eq:3.52}
\end{equation}
Here, the stability condition of the solution (\ref{eq:3.52}) is given from Eq.~(\ref{eq:3.51}) by $a_1\cos2\psi<0$. 
Then, by using Eq.~(\ref{eq:3.52}), the condition (\ref{eq:3.50}) for $v=0$ becomes $\gamma < \gamma^{\dag}$, where
\begin{equation}
\gamma^{\dag} = \frac{a_1}{2} \left( K - \frac{4 (b_2-a_2)^2 \tilde{\kappa} }{a_1^2} \right)^{1/2}  .
 \label{eq:3.53}
\end{equation}
This expression is valid as long as Eq.~(\ref{eq:3.52}) satisfies  $|\sin 2\psi|\le1$ which imposes a constraint for $b_3$
 as $b_3\ge b_3^{\dag}$, where
\begin{equation}
b_3^{\dag} =  \frac{ \kappa }{\zeta} \left\{ 1+ \frac{4 c_2 (b_2-a_2)^2 }{a_1^2} \right\}  .
 \label{eq:3.54}
\end{equation} 
On the other hand, for $b_3^*<b_3<b_3^{\dag}$, Eq.~(\ref{eq:3.51}) has no steady solution, and hence $\psi$ varies from 0 to $\pi$. 
Therefore, on the average of $\cos 2\psi$, the condition for $v=0$, given by Eq.~(\ref{eq:3.50}), becomes $\gamma<0$. 
Consequently, the stability threshold of $v=0$ in Eq.~(\ref{eq:3.1}) is given by $\gamma=0$ for $0<b_3<b_3^{\dag}$ and $\gamma=
\gamma^{\dag}$ for $b_3>b_3^{\dag}$ where $b_3^{\dag} =0.375$. 
The bifurcation threshold from $v=0$ to a finite $v$ obtained in this way is shown by the thin solid line in Fig.~\ref{fig:diagram}, which is consistent  with the numerical results of Eqs.~(\ref{eq:1.7}) - (\ref{eq:1.11}).

Next, we derive the stability condition for the migrating states $v\ne 0$. The velocity is given from Eq.~(\ref{eq:3.1}) by $v=\sqrt{\Gamma}$ with $\Gamma$ given by Eq.~(\ref{eq:3.2}). 
Then, from Eqs.~(\ref{eq:1.8}) - (\ref{eq:1.10}), the time-evolution equations become  
\begin{eqnarray}
\frac{d s}{d t} &=& -\kappa(1+B \cos^2 2\psi) s +b_1 \gamma \cos2\psi +b_3 s \omega^2   ,
 \label{eq:3.16}\\
\frac{d \psi}{d t} &=& -\left( \frac{b_1\gamma}{2s}-\frac{\kappa}{2} B \cos2\psi -\frac{a_1}{2} s \right) \sin2\psi  
 + ( b_2-a_2 ) \omega  ,
 \label{eq:3.18}
\end{eqnarray}
where $\omega$ and $B$ are given by Eq.~(\ref{eq:3.19}) and 
 Eq.~(\ref{eq:B}), respectively. 
When $\omega=\pm\tilde{\omega}$,  there is a constraint $\zeta-c_2 s^2>0$ as well as $s>0$ and $\Gamma>0$.
The linear stability matrix of Eqs.~(\ref{eq:3.16}) and (\ref{eq:3.18})  is defined by 
\begin{equation}
L(s,\psi) = 
 \left[
 \begin{array}{cc}
 \partial_s (ds/dt) & \partial_{\psi} (ds/dt) \\
 \partial_s (d\psi/dt) & \partial_{\psi} (d\psi/dt)
 \end{array}
 \right]    ,
 \label{eq:3.27}
\end{equation}
where $\partial_{s}$ means the partial derivative with respect to $s$. 
The method we employ to obtain the stability of the fixed points is as follows; 
first, 
we evaluate numerically  the fixed point of  Eqs.~(\ref{eq:3.16}) and (\ref{eq:3.18}) with $\omega=\pm\tilde{\omega}$ under the constraints $\Gamma>0$ and $\zeta-c_2s^2>0$. 
Then, by using the determinant and the trace of the linear stability matrix (\ref{eq:3.27}), we can examine the stability of the fixed point. 
Note that the reduced equations (\ref{eq:3.16}) and (\ref{eq:3.18}) are invariant under the simultaneous transition $\omega\to-\omega$ and $\psi\to\pi-\psi$. 
Therefore, we will consider only the case of $\omega=+\tilde{\omega}$ hereafter.

The nullclines and the fixed points obtained from Eqs.~(\ref{eq:3.16}) and (\ref{eq:3.18}) with $\omega=\tilde{\omega}$ 
are plotted on the $s$-$\psi$ plane as shown in Fig.~\ref{fig:s_psi}, where the attractors obtained numerically from Eqs.~(\ref{eq:1.7}) - 
(\ref{eq:1.11}) are also superposed. 
The $s$- and $\psi$-nullclines are displayed by the broken line and the dotted line respectively. 
The black squares and the white circles indicate the saddle points and the unstable fixed points of Eqs.~(\ref{eq:3.16}) and (\ref{eq:3.18}). 
The double circles and the black circles 
are the attractors of the revolution I and II states respectively obtained from Eqs.~(\ref{eq:1.7}) - (\ref{eq:1.11}), which agree, within numerical accuracy, with the stable fixed points obtained from  Eqs.~(\ref{eq:3.16}) and (\ref{eq:3.18}). The limit cycle orbit corresponding to the quasi-periodic I state obtained from Eqs.~(\ref{eq:1.7}) - (\ref{eq:1.11}) exists around the unstable fixed point. 
The stability analysis of the fixed points of
 the reduced equations,  Eqs.~(\ref{eq:3.16}) and (\ref{eq:3.18}) with $\omega=\tilde{\omega}$, show a good agreement with the bifurcation behavior  
 obtained from the numerical simulations of the original equations (\ref{eq:1.7}) - (\ref{eq:1.11}).

\begin{figure}[tb]
  \begin{center}
    \includegraphics[width=10cm]{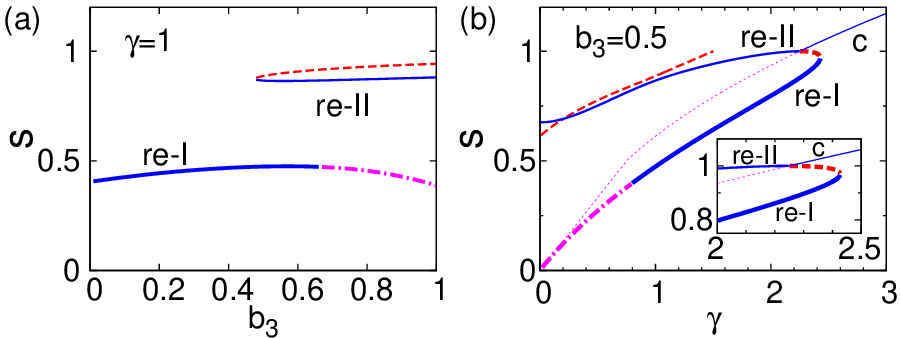} 
      \caption{(color online) 
     Bifurcation diagram varying (a) $b_3$ for $\gamma=1$ and (b) $\gamma$ for $b_3=0.5$. 
      The stationary value of  $s$ is obtained from the reduced equations (\ref{eq:3.16}) and (\ref{eq:3.18}) with $\omega$ given by Eq.~(\ref{eq:3.19}). 
      The thick and thin solid lines indicate revolution I and II states respectively. The dotted broken line and the dotted line indicate the unstable solutions.
      In Fig.~(b), the stable solution  with $\omega=0$, which corresponds to the circular state, is also shown (the line indicated by ``c").
      } \label{fig:s_bif}
  \end{center}
\end{figure}

The stable and unstable blanches  of the steady solution $s$ for $\omega \ne 0$ are displayed in Fig.~\ref{fig:s_bif}(a) for $\gamma=1$ and by varying the value of $b_3$.
There are at most three fixed points; 
One is plotted by the thick solid line, which corresponds to the stable revolution I state. 
This fixed point loses its stability at around $b_3=0.66$ by a Hopf bifurcation and becomes an unstable fixed point, displayed by the thick dotted-broken line. 
At around $b_3=0.48$, a pair of another stable fixed point and a saddle point appears by a saddle node bifurcation, which are displayed by the thin solid line and the  broken line, respectively. 
This stable fixed point corresponds to the revolution II state.

Finally, we investigate the dynamics around the region $\omega=0$. 
The fixed point of Eqs.~(\ref{eq:3.16}) and (\ref{eq:3.18}) with $\omega=0$ is given by 
\begin{equation}
s = \frac{b_1 \gamma \cos 2\psi}{\kappa \left( 1+B\cos^2 2\psi \right)}  ,
 \label{eq:3.22}
\end{equation}
and 
\begin{equation}
\tan 2\psi  = 0  
 \begin{array}{cc}
 \rm{~~~~~~~~~~~~~~~~~~for~} \gamma \le \gamma_c     ,
  \end{array}  
  \end{equation}
  \begin{equation}
\cos^2 2\psi = \frac{\kappa}{2B (\gamma-\kappa/2)}
   \begin{array}{cc}
\rm{~~for~} \gamma \ge \gamma_c    ,
 \end{array}
 \label{eq:3.24}
\end{equation}
where $\gamma_c$ has been defined by Eq.~(\ref{eq:gamma_c}).  
From Eq.~(\ref{eq:3.22}), the requirement of the positivity of $v$ in Eq.~(\ref{eq:3.1})  gives us 
\begin{equation}
\Gamma = \gamma -\frac{B \gamma \cos^2 2\psi}{1+B\cos^2 2\psi} 
 = \frac{\gamma}{1 +B\cos^2 2\psi} > 0   .
 \label{eq:3.38}
\end{equation}
Since $B$ is positive from the definition (\ref{eq:B}), this requirement is satisfied as long as $\gamma > 0$. 
These results are consistent with those in Ref.~\cite{OhtaOhkuma}, where the spinning motion was not considered.
The condition  $\zeta -c_2 s^2 \le 0$ must be satisfied for  $\omega=0$ in Eq.~(\ref{eq:3.19}).
From Eqs.~(\ref{eq:3.22}) and (\ref{eq:3.24}), this condition is written as 
\begin{equation}
\zeta - \frac{c_2 (b_1 \gamma)^2 \cos^2 2\psi}{\kappa^2 (1+B \cos^2 2\psi)^2} 
 = \zeta - \frac{c_2 b_1^2 (\gamma -\kappa/2) }{2 \kappa B} 
 \le 0  .
 \label{eq:3.36}
\end{equation}
This leads to $\gamma\ge\gamma^*$ for $c_2>0$ where  the bifurcation threshold $\gamma^*$ is given by 
\begin{equation}
\gamma^* = \frac{\kappa}{2} + \frac{2 \kappa B \zeta}{b_1^2 c_2} = \frac{\kappa}{2} + \frac{a_1 \zeta}{b_1 c_2}  .
 \label{eq:3.37}
\end{equation}
It is noted from Eqs.~(\ref{eq:gamma_c}) and (\ref{eq:3.37}) that $\gamma_c=0.75$ and $\gamma^*=2.25$.

\begin{figure}[tb]
  \begin{center}
   \includegraphics[width=10cm]{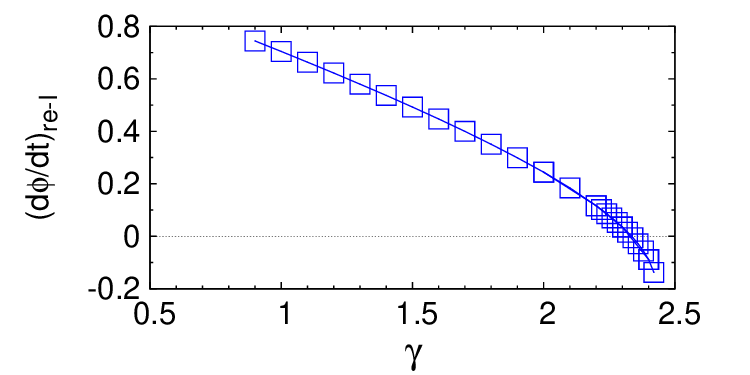} 
      \caption{(color online) 
      Angular velocity $d\phi/dt$, given by Eq.~(\ref{eq:1.8}), of the revolution I state with $\omega>0$ obtained numerically from Eqs.~(\ref{eq:1.7}) - (\ref{eq:1.11}). 
      Between $\gamma=2.33$ and $2.34$, the sign of $d\phi/dt$ changes from positive to negative, where the radius of the circular trajectory in real space diverges as shown in Fig.~\ref{fig:trace_c}(d). 
      } \label{fig:c_phi}
  \end{center}
\end{figure}

These results are summarized in Fig.~\ref{fig:s_bif}(b), where  the fixed point values of $s$ and its stability are plotted for $\omega=0$  together with those for $\omega=\tilde{\omega}$ by changing the values of $\gamma$ for the fixed value of $b_3=0.5$. 
For large values of $\gamma$, there is only one stable fixed point with $\omega=0$ shown by the thin solid line in Fig.~\ref{fig:s_bif}(b), which corresponds to the circular state. 
Around $\gamma=2.427$, another stable fixed point and an unstable saddle point  with finite $\omega$ appear by a saddle node bifurcation,  which are shown by the thick solid line (re-I) and by the thick broken line, respectively. 
By decreasing the value of  $\gamma$, the stable fixed point, which corresponds to the revolution I state, becomes unstable around $\gamma=0.8$. On the other hand, the line of the saddle point crosses with the line of the stable fixed point of the circular state at around $\gamma=\gamma^*$. 
Then, the circular state loses its stability and the saddle point become stable by a transcritical bifurcation. 
This stable fixed point indicated by the solid line (re-II) corresponds to the revolution II state. 

It is mentioned that the anomalous behavior of the radius in the revolution  I motion at  $\gamma \approx 2.33$ 
shown in Fig.~\ref{fig:trace_c}(d)
is reproduced by the reduced set of equations  (\ref{eq:3.16}) and (\ref{eq:3.18}) with Eq.~(\ref{eq:1.8}). 
In Fig.~\ref{fig:c_phi}, we have plotted $d\phi/dt$ as a function of $\gamma$ substituting the stationary values  of the revolution I state.
It is found that the angular velocity changes its sign around $\gamma \approx 2.33$   from positive to negative by increasing $\gamma$. This causes an anomalously large radius of the circular trajectory in the real space and the change of rotating direction.

In summary,  the reduced equations (\ref{eq:3.16}) and (\ref{eq:3.18}) for $s$ and $\psi$, and the time-evolution equation (\ref{eq:1.8}) for $\phi$ together with $v=\sqrt{\Gamma}$ and $\omega=\pm\tilde{\omega}$ reproduce all of the revolution I and II states, the circular state and the quasi-periodic I and II states. However, they do not reproduce the periodic-doubling state and the chaotic state.

\section{Discussion}\label{sec:discussion}

In this paper, we have investigated in two dimensions  the dynamics of a deformable self-propelled particle having a spinning degree of freedom. 
By numerical simulations of Eqs.  (\ref{eq:1.1}) - (\ref{eq:1.3}), we have obtained a dynamical phase diagram as shown in Fig.~\ref{fig:diagram} for $\kappa=0.5$, which displays a rich variety of dynamical states.
Apart from the motionless state, there are the spinning state, three types of revolution states and two quasi-periodic states. 
The trajectories in the real space of the latter five states are displayed in Fig.~\ref{fig:trace_c}, Fig.~\ref{fig:qp1}(a) and Fig.~\ref{fig:qp2}(a). 
The period-doubling and chaotic motions, as shown in Fig.~\ref{fig:trace_pd}, are also obtained between the quasi-periodic I and II states.

Theoretical analysis has been developed for  the bifurcations between these dynamical states based on the two-variable equations (\ref{eq:3.3}) and (\ref{eq:3.4}) for $v=0$, and Eqs.~(\ref{eq:3.16}) and (\ref{eq:3.18}) for finite values of $v$.  This simplified set of equations succeeds in reproducing all the bifurcations except for the chaotic behavior between 
the quasi-periodic I and II states. Although not described here, we have verified that all of the  dynamics can be reproduced if we employ the three-variable system in terms of $s$, $\psi$, and $\omega$. However, another set of three variables $s$, $\psi$, and $v$ is unable to realize the chaotic state.

We have found numerically an anomalous increase of the radius in the revolution I state as shown in Fig.~\ref{fig:trace_c}(d). 
  This anomaly originates from the existence of zero in  $d\phi/dt$ as a function of $\gamma$ in the revolution I state as shown in Fig.~\ref{fig:c_phi}. This is obtained numerically from the original equations (\ref{eq:1.7}) - (\ref{eq:1.11}) for $b_3=0.5$ with $\omega>0$. The angular velocity $d\phi/dt$ is positive for $\gamma\lesssim2.33$ while is becomes negative for $\gamma\gtrsim2.34$.  Therefore, a particle in the revolution I state with $\omega>0$ undergoes a counter-clockwise orbital rotation for $\gamma\lesssim2.33$, whereas it rotates to the clockwise direction for $\gamma\gtrsim2.34$. This implies that if a noise term is added in Eq.~(\ref{eq:1.8}), a particle changes randomly the rotation direction in the vicinity of the anomalous point.

Our representation of the basic set of equations  (\ref{eq:1.1}) - (\ref{eq:1.4}) is independent of the dimensionality. Therefore, the present study of spinning motion of a soft particle can be readily extended to three dimensions. It is of particular interest to investigate the case that the spinning axis is parallel to the migration velocity. This corresponds to the motion of flagellated bacteria \cite{DiLuzio, Goldstein}. We will return to this problem somewhere in the near future.

\section*{Acknowledgments}

MT would like  to  thank to Prof. H. Brand for valuable discussion. 
This work was supported by the JSPS Core-to-Core Program ``International research network for non-equilibrium dynamics of soft matter" and 
by Grant-in-Aid for Scientific Research C (No. 23540449) from JSPS.

\end{document}